\documentclass[preprint]{aastex}

\begin{document}

\title{The Light and Period Variations of the Eclipsing Binary AA Ursae Majoris}
\author{Jae Woo Lee, Chung-Uk Lee, Seung-Lee Kim, Ho-Il Kim, and Jang-Ho Park}
\affil{Korea Astronomy and Space Science Institute, Daejeon 305-348, Korea}
\email{jwlee@kasi.re.kr, leecu@kasi.re.kr, slkim@kasi.re.kr, hikim@kasi.re.kr, pooh107162@kasi.re.kr}

\begin{abstract}
We present new multiband CCD photometry for AA UMa made on 8 nights between January and March 2009; the $R$ light curves are 
the first ever compiled. Historical light curves, as well as ours, display partial eclipses and inverse O'Connell effects 
with Max I fainter than Max II.  Among possible spot models, a cool spot on either of the component stars and its variability 
with time permit good light-curve representations for the system. A total of 194 eclipse timings over 81 yrs, including 
our five timings, were used for ephemeris computations. We found that the orbital period of the system has varied due to 
a periodic oscillation overlaid on an upward parabolic variation.  The continuous period increase at a fractional rate of 
$+$1.3$\times$10$^{-10}$ is consistent with that calculated from the W-D code and can be interpreted as a thermal mass transfer 
from the less to the more massive secondary star at a rate of 6.6$\times$10$^{-8}$ M$_\odot$ yr$^{-1}$. The periodic component 
is in satisfactory accord with a light-time effect due to an unseen companion with a period of 28.2 yrs, a semi-amplitude 
of 0.007 d, and a minimum mass of $M_3 \sin i_3$=0.25 $M_\odot$ but this period variation could also arise from magnetic activity. 
\end{abstract}

{}

\section{INTRODUCTION}

The low-temperature contact binary AA UMa (GSC 3433-0685, TYC 3433-685-1, 2MASS J09465929+4545563; $V$=$+$10.97, G0 V) was 
discovered to be an eclipsing variable by Hoffmeister (1948).  Meinunger (1976) and Lu et al. (1988) made 
photoelectric light curves in the $UBV$ and $BV$ bandpasses, respectively.  Lu (1988) obtained double-line radial velocity curves 
with values of $K_1$=124.0 km s$^{-1}$ and $K_2$=227.7 km s$^{-1}$. He classified the spectral types of both components to be 
around G0 V, in good agreement with the color index $B-V$=$+$0.60 at maximum light given by Lu et al. Simultaneous analyses of 
these radial velocities and the Lu et al. light curves were performed by Wang \& Lu (1990) and Barone et al. (1993). 
Their two solutions agree with each other within the limits of the errors for most parameters, but the former results indicated 
that the system is a W-subtype (defined observationally by Binnendijk 1970) contact binary with a fill-out factor of $f$=15 \% 
while the latter led to a nearly contact system with respect to the critical Roche lobes. Finally, Qian (2001) reported that 
the orbital period is secularly increasing at a rate of +9.05$\times$10$^{-8}$ d yr$^{-1}$. In this paper, we present and discuss 
the physical properties of the eclipsing system from detailed studies of the historical and new eclipse timings and light curves.

\section{OBSERVATIONS}

AA UMa was observed on eight nights from 2009 January 8 through March 8, using a SITe 2K CCD camera and a $BVR$ filter set 
attached to the 61-cm reflector at Sobaeksan Optical Astronomy Observatory (SOAO) in Korea. The instrument and reduction method 
have been described by Lee et al. (2007b).  In order to look for a comparison star suitable for AA UMa and for probably 
variable objects, we monitored a few tens of field stars imaged on the chip at the same time as the program target. 
As in the process described by Lee et al. (2010), we made an artificial reference source from several stars on the CCD frame 
and examined candidate comparison stars.  By this procedure, GSC 3433-0352 (TYC 3433-352-1; $V$=$+$11.1, $B-V$=$+$0.80) was 
chosen as an optimal comparison ($C$) star.  An observed image is given as Figure 1, wherein $N1$ (SDSS J094743.02+455529.5) 
and $N2$ (SDSS J094558.17+454814.2) were discovered to be new eclipsing binaries with tentative periods of about 1.5517 d 
and 1.9345 d, respectively. In the near future, the light curves of the two systems will be presented in a separate paper. 
A total of 893 individual observations were obtained in the three bandpasses (299 in $B$, 304 in $V$, and 290 in $R$) and 
a sample of them is listed in Table 1. The natural-system light curves of AA UMa are plotted in Figure 2 as 
differential magnitudes {\it versus} orbital phase. The $R$ light curve is the first ever compiled.

\section{LIGHT-CURVE SYNTHESIS AND SPOT MODEL}

Light curves of AA UMa are typical of a short period contact binary, and the curved bottoms of both minima indicate 
partial eclipses. As do the light curves of Lu et al. (1988), our SOAO observations display the O'Connell effect with 
Max I fainter than Max II by about 0.054, 0.043, and 0.035 mag for the $B$, $V$, and $R$ bandpasses, respectively. 
The light curves were analyzed in a manner similar to that for the cool contact systems BX Peg (Lee et al. 2004) and 
TU Boo (Lee et al. 2007a).  Contact mode 3 of the 2003 version\footnote {ftp://ftp.astro.ufl.edu/pub/wilson/} of 
the Wilson-Devinney code (Wilson \& Devinney 1971; hereafter W-D) was applied to the light curves normalized to 
unit light at phase 0.75. The effective temperature ($T_2$) of the more massive star was fixed at 5920 K, 
according to the spectral type G0 V and intrinsic color index $(B-V)\rm_{0}$=$+$0.59 after assuming $E$($B-V$)=$+$0.01 
(Schlegel et al. 1998). The logarithmic bolometric ($X$, $Y$) and monochromatic ($x$, $y$) limb-darkening coefficients 
were interpolated from the values of van Hamme (1993) in concert with the model atmosphere option. The initial value 
for a mass ratio ($q=m\rm_2/m\rm_1$) was taken from Wang \& Lu (1990). A third light source ($\ell_3$) was considered 
throughout the analysis but remained zero within its error.  In Table 2, the parameters with parenthesized formal errors 
signify adjusted ones and the subscripts 1 and 2 refer to the primary and secondary stars being eclipsed at Min I and 
Min II, respectively. Thus, the latter star is the cooler, larger, and more massive component. 

The unspotted solution for the SOAO data is listed in the second column of Table 2 and appears as the dashed curves 
in Figure 2. The residuals from this model are plotted in the leftmost panels of Figure 3, wherein it can be seen that 
these initial model light curves do not fit the observed ones at all near phase 0.25. It has been accepted that 
intrinsic stellar activity causes the asymmetries in the light maxima of many other contact binaries on our program, 
such as TU Boo (Lee et al. 2007a), BX Peg (Lee et al. 2009a), AR Boo (Lee et al. 2009b), and GW Cep (Lee et al. 2010). 
In principle, the asymmetries might be due to large cool starspots, to hot regions such as faculae, or to gas streams 
and their impact on a companion star. Because there is, at present, no {\it a priori} way to know which phenomenon is 
more appropriate to explain the asymmetric light curves, we reanalyzed our light curves testing each of a hot and 
cool spot on each component.  The converged results are given in columns (3)--(6) of Table 2 together with 
the spot parameters. The residuals from the spot models are plotted in the second to fifth multiple panels of Figure 3.  
We see that the cool-spot models (Cool 1 and Cool 2) give very slightly smaller values for the sum of the residuals 
squared ($\Sigma W(O-C)^2$) than do the hot-spot models (Hot 1 and Hot 2). The model light curves from Cool 2 are 
plotted as the solid curves in Figure 2 and represent the observed asymmetries satisfactorily although it must be noted 
that there are unexplained bifurcations in the pattern of residuals in the primary eclipse.

To study the spot behavior of AA UMa further and to examine whether our solutions can reasonably describe 
the light curves of Lu et al. (1988), we analyzed the old light curves for the two cases of a cool spot on 
each component star. For this test, we used our binary parameters as initial values.  The results appear in Table 3 
and are plotted in Figure 4 as the continuous curves. The parameters from the Lu et al. light curves are consistent 
with those from our data except for the orbital inclination, i.e., the earlier value is about 0$^\circ$.8 larger 
than the later one. We can offer no explanation for this discrepancy. From his spectra, Lu (1988) noted that 
some distortions appeared in the cross correlation profiles for AA UMa and suggested that these features can be 
originated in stellar spots, rather than from sporadic flare eruptions on the more massive secondary component. 
Long ago, Mullan (1975) suggested that the more massive components of contact binaries would preferentially 
manifest magnetic starspots.  It would be useful to treat Meinunger's (1976) data in the same way but they are 
not currently available.

In reality, it is not easy to distinguish between the spot models because the differences among them are so small 
and because the two cool spot models are really degenerate:  no matter which model is chosen, the spot must be 
turned to the observer at phase 0.25.  A more detailed model conceives of a cool spot on each component star at 
the stellar coordinates already evaluated.  The spot radii and temperature ratios naturally change from the values 
in Table 2 but the fit is not so good as when a single spot is postulated for a single star.  Following Mullan, 
then, we believe that a large, dark spot on the secondary star and its variability with time provide a model testable 
for the near future.

\section{ORBITAL PERIOD STUDY}

From our observations, five weighted times of minimum light were determined by using the method of Kwee \& van Woerden (1956). 
In addition to these, 194 eclipse timings (41 photographic plate, 79 visual, 7 photographic, 27 photoelectric and 40 CCD) 
have been collected from the data base of Kreiner et al. (2001) and from more recent literature. All available photoelectric 
and CCD timings are listed in Table 4.  For the period analysis of AA UMa, because many timings of the system have been published 
with no errors, the following standard deviations were assigned to timing residuals based on observational method: $\pm$0.068 d 
for photographic plate, $\pm$0.008 d for visual, $\pm$0.017 d for photographic, and $\pm$0.001 d for photoelectric and CCD minima. 
Relative weights were then scaled from the inverse squares of these values (Lee et al. 2007b).

The period variation of the system was studied for the first time by Qian (2001). Based on only 15 photoelectric times 
of minimum light, he reported that there exists a parabolic variation, indicating a continuous period increase. To test 
this possibility, we fitted all times of minimum light to his quadratic ephemeris but this failed to do justice to 
all subsequent timings because the $O$--$C$ residuals display a sine-like variation superposed on an upward parabola. 
The periodic component suggests a light-travel time (LTT) effect caused by the presence of a third body in the system.  
Thus, all eclipse timing residuals were fitted to a quadratic {\it plus} LTT ephemeris:
\begin{eqnarray}
C = T_0 + PE + AE^2 + \tau_{3},
\end{eqnarray}
where $\tau_{3}$ is the LTT due to an additional distant companion orbiting the eclipsing pair (Irwin 1952) and includes 
five parameters ($a_{12}\sin i_3$, $e$, $\omega$, $n$ and $T$).  The Levenberg-Marquardt algorithm (Press et al. 1992) 
was applied to solve for the eight parameters of the ephemeris (Irwin 1959) and the results are summarized in Table 5, 
together with related quantities.  The absolute dimensions presented in a later section have been used for these and 
subsequent calculations.  The $O$--$C$ diagram constructed with the linear terms of the quadratic {\it plus} LTT ephemeris 
is plotted in the top panel of Figure 5, where the continuous curve and the dashed parabola represent the full contribution 
and the quadratic term, respectively. The middle panel displays the LTT orbit, and the bottom panel the PE and CCD residuals 
from the complete ephemeris. These appear as $O$--$C_{\rm full}$ in the third column of Table 4. As displayed in the figure, 
all modern times of minimum light currently agree with a quadratic {\it plus} LTT ephemeris quite well. If the third object 
is on the main sequence and its orbit is coplanar with the eclipsing binary ($i_3 \simeq$ 80$^\circ$), the mass and radius of 
the object are computed to be $M_3$ = 0.25 M$_\odot$ and $R_3$ = 0.27 R$_\odot$, following the empirical mass-radius relation 
of Bayless \& Orosz (2006). Because these values would correspond to a spectral type of M5 and a bolometric luminosity of 
$L_3$ = 0.007 L$_\odot$, it will be difficult to detect such a companion from analyses of spotted-star light curves and 
spectroscopic observations but the absence of this evidence cannot rule out the existence of a companion orbiting the close pair.

The periodic variation could possibly be produced by asymmetrical eclipse minima due to stellar activity and/or even by 
the method of measuring the minimum epochs (Maceroni \& van't Veer 1994, Lee et al. 2009b). The light curve synthesis method 
gives more precise timings than do other techniques which do not consider spot activity and are based only on observations 
during eclipsing minima. Because all available light curves were modeled for spot parameters, we determined 
nine light curve timings with the W-D code by means of adjusting only the ephemeris epoch ($T_0$) in Tables 2 and 3. 
The results are listed in the second column of Table 6 together with the previously-tabulated timings and are illustrated 
by the plus symbols in Figure 5. We can see that the differences among them are much smaller than the observed semi-amplitude 
(about 0.007 d) of the cyclical variation and that the W-D timings agree with our analysis of the $O$--$C$ diagram. So, 
the orbital period of AA UMa has really varied due to a periodic oscillation plus an upward parabola. There definitely exist 
systematic runs of differences between the observed timings and the calculated W-D ones.  These epoch differences are positive 
for Min I while those for Min II are negative because a cool spot seen at the first quadrature produces positive values 
for Min I and negative values for Min II.

\section{DISCUSSION}

It should be recognized that our light curve representation is closer to Wang \& Lu's (1990) in calling for 
an over-contact system than it is to that of Barone et al. (1993) and that a cool spot demands a fillout factor larger 
than does a hot spot.  We analyzed the velocity curves of Lu (1988) with our photometric solutions for Cool 2 and 
computed the astrophysical parameters for the system listed in Table 7, together with those of Wang \& Lu for comparison.  
Agreement is good.  The luminosity ($L$) and the bolometric magnitudes ($M_{\rm bol}$) were obtained by adopting 
$T_{\rm eff}$$_\odot$=5780 K and $M_{\rm bol}$$_\odot$=+4.69 for solar values (Popper 1980).  The temperature of 
each component was assumed to have an error of 150 K and the bolometric corrections (BCs) were obtained from the relation 
between $\log T_{\rm eff}$ and BC given by Kang et al. (2007). Using apparent visual magnitude of $V$=+10.97 at Max II 
(Lu et al. 1988) and interstellar reddening of $A_{\rm V}$=0.03, we have calculated the distance of the system 
to be 350 pc. In order to consider the evolutionary status of AA UMa, our absolute parameters were compared with 
the mass-radius, mass-luminosity and HR diagrams from Hilditch et al. (1988). In these diagrams, both components 
do conform to the general pattern of contact binaries.  

The coefficient (A) of the quadratic term in equation (1) is positive and signifies a secular period increase with 
a rate of d$P$/d$t$ = $+$(4.7$\pm$0.7) $\times 10^{-8}$ d yr$^{-1}$, corresponding to a fractional period change of 
$+$(1.3$\pm$0.2)$\times$10$^{-10}$. Within errors, this value is close to $+$(4.8$\pm$1.5)$\times$10$^{-10}$ derived 
from our W-D synthesis of all AA UMa curves (Lu 1988, Lu et al. 1988, SOAO data) obtained over about 23 yrs, independently 
of the eclipse timings. Under the assumption of conservative mass transfer, the smaller, more precise rate gives 
a continuous mass transfer from the less massive primary to the more massive secondary star at the modest rate of 
6.6$\times$10$^{-8}$ M$_\odot$ yr$^{-1}$.  This agrees well with the rate of 6.5$\times$10$^{-8}$ M$_\odot$ yr$^{-1}$ 
calculated by assuming that the primary transfers its present mass to the secondary on a thermal time scale. 
According to the thermal relaxation oscillation (TRO) theory (Lucy 1976; Lucy \& Wilson 1979), contact binaries must 
oscillate cyclically between contact and non-contact conditions. Because our detailed study of AA UMa indicates that 
the orbital period is increasing and that mass is moving between the component stars, the system may presently be in 
an expanding TRO state evolving from a contact to a non-contact condition as the theory permits.

Because the LTT hypothesis does not produce a unique explanation, we also consider the possibility that 
the periodic component of the $O$--$C$ residuals could be produced by a magnetic activity cycle 
(Applegate 1992, Lanza et al. 1998). According to this mechanism, a variation in the distribution of angular momentum 
of a magnetically active star produces a change in the star's gravitational quadratic moment and hence forces a modulation 
of the orbital period. In order to explain period modulations of $\Delta P/P \sim 10^{-5}$, this model typically requires 
that the active star be variable at the $\Delta L/L \simeq 0.1$ level and differentially rotating at 
the $\Delta \Omega/\Omega \simeq 0.01$ level. In W UMa-type contact binaries, the amplitudes of the modulations are 
usually of the order of $\Delta P/P \sim 10^{-6}$ and  $\Delta \Omega/\Omega \sim$ 0.001$-$0.003. This would be sufficient 
to account for the period variation. With the period ($P_3$) and amplitude ($K$) listed in Table 5, the model parameters 
for AA UMa were calculated from the formulae given by Applegate and are listed in Table 8, where the rms luminosity changes 
($\Delta m_{\rm rms}$) converted to magnitude scale were computed with the equation (4) of Kim et al. (1997). 
The tabulated results correspond to typical values for contact binaries (cf. Lanza \& Rodono 1999) and indicate that 
the Applegate mechanism could possibly function in both stars. Observationally, the Applegate model predicts that 
orbital period changes should be consistent with light and color variations occuring in the same cycle. However, we cannot 
check this possibility because there are only 2 epochs of light curves so far. More systematic and continuous photometric 
monitoring will help to understand the orbital period change and the long-term spot behavior of AA UMa.

\acknowledgments{ }
The authors wish to thank Dr. Robert H. Koch for careful readings and corrections and for some helpful comments on 
the draft version of the manuscript. We also thank the staff of the Sobaeksan Optical Astronomy Observatory for assistance 
with our observations. This research has made use of the Simbad database maintained at CDS, Strasbourg, France.

\clearpage
\begin{figure}
 \includegraphics[]{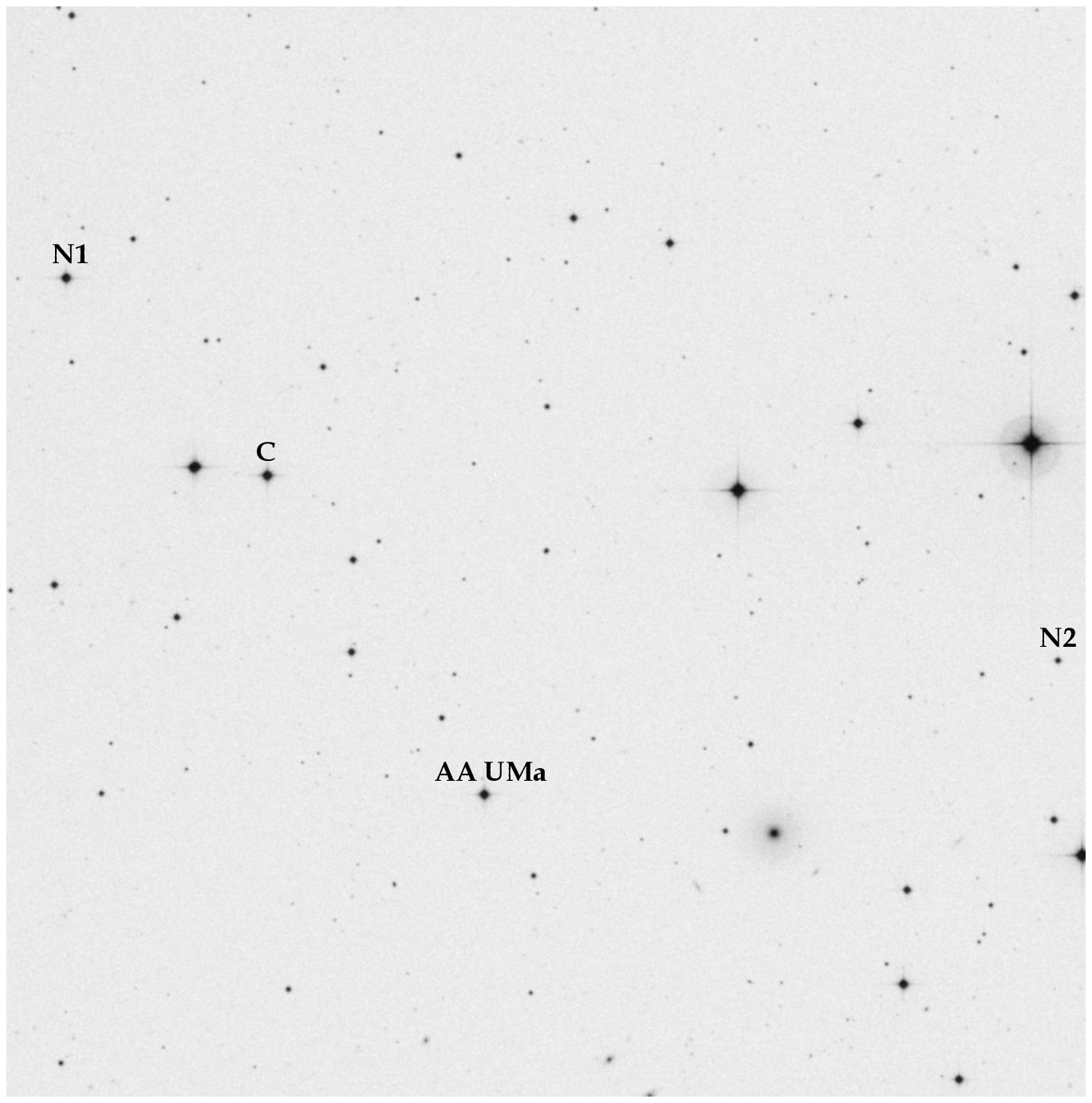}
 \caption{An observed CCD image (20$\arcmin$.5$\times$20\arcmin.5) of AA UMa and many nearby stars. $N1$ and $N2$ 
 are eclipsing binary systems newly discovered from our measurements.  Monitoring numerous frames led us to choose 
 the star $C$ as a comparison.  North is up and east is to the left. }
 \label{Fig1}
\end{figure}

\begin{figure}
 \includegraphics[]{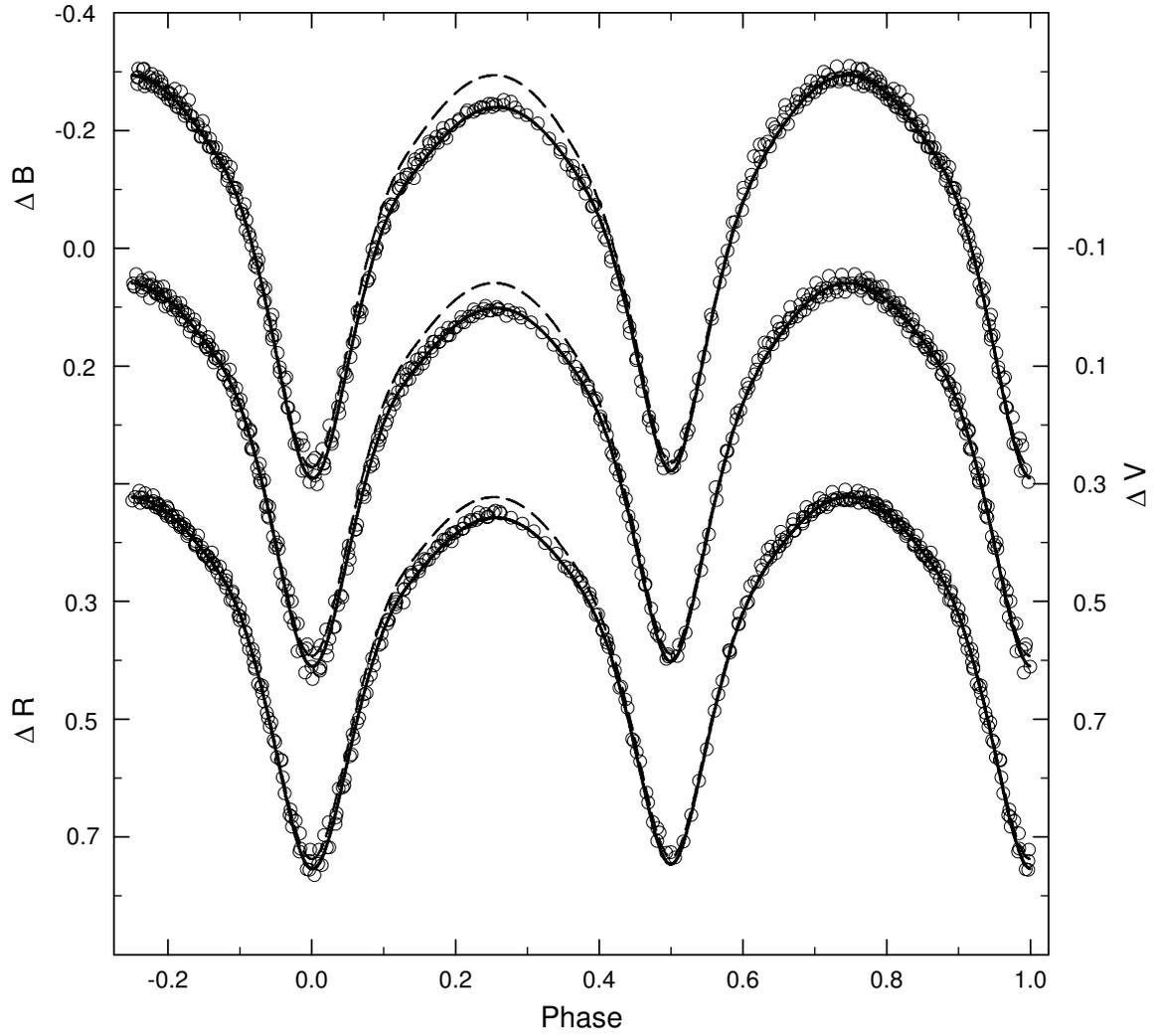}
  \caption{$BVR$ observations of AA UMa with the fitted model light curves.  The dashed and solid curves are computed 
  without and with a spot, respectively. }
 \label{Fig2}
\end{figure}

\begin{figure}
 \includegraphics[]{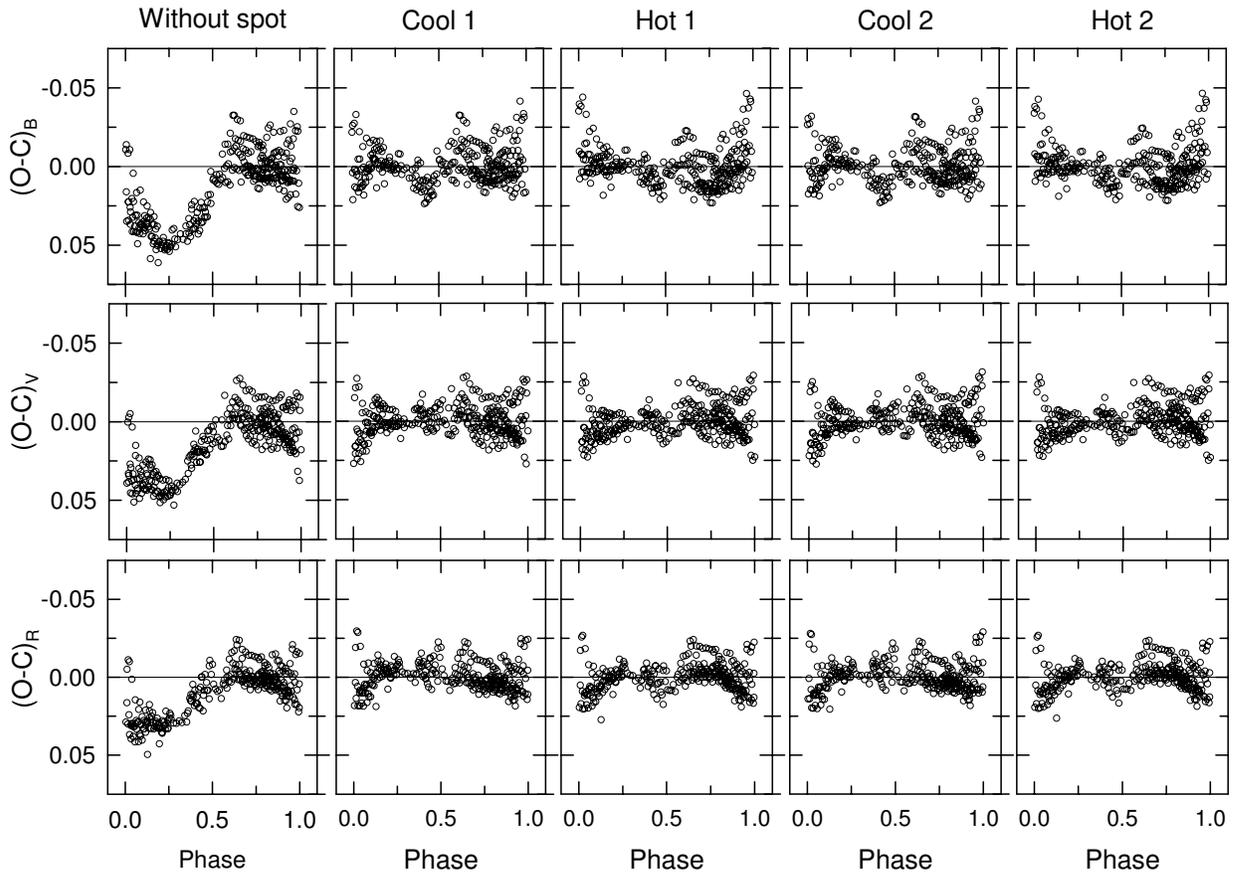}
 \caption{Light residuals in the $B$, $V$, and $R$ bandpasses corresponding to our binary models listed in Table 2. }
 \label{Fig3}
\end{figure}

\begin{figure}
 \includegraphics[]{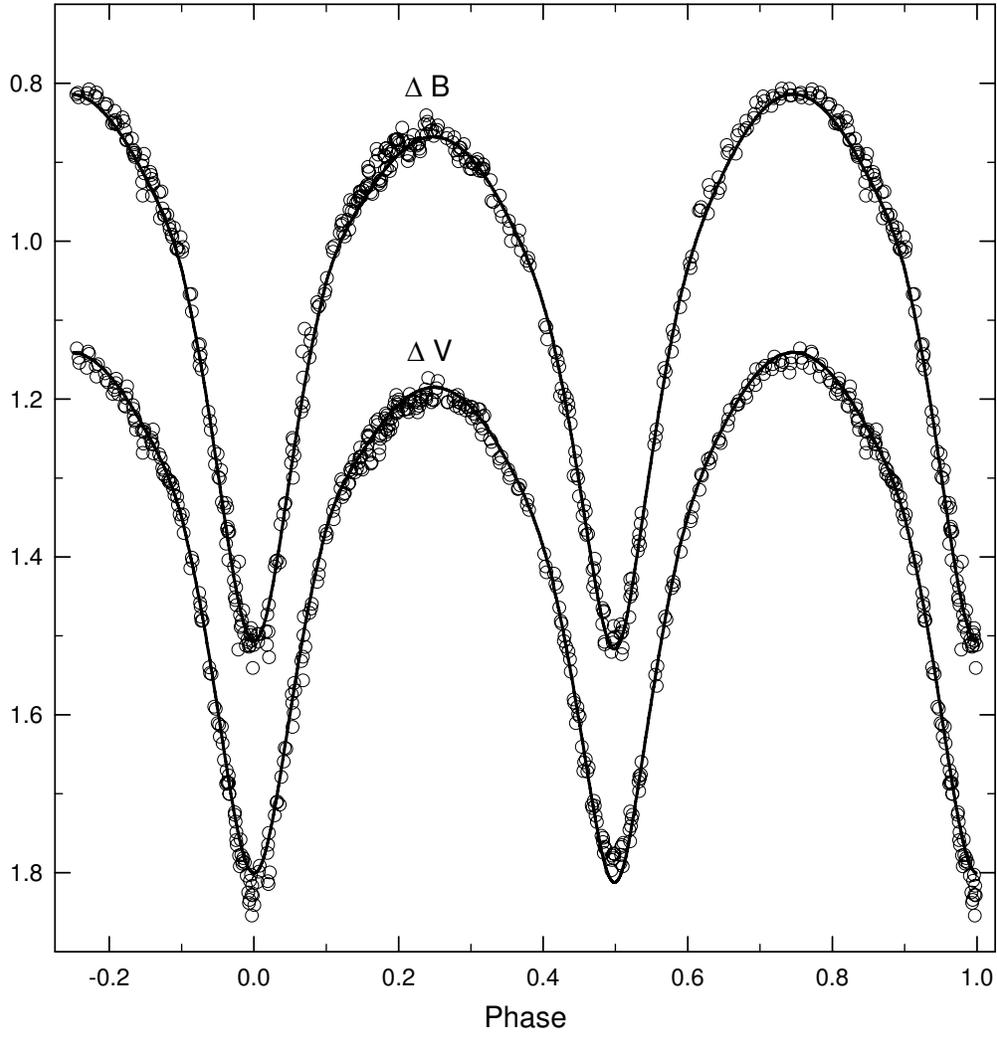}
 \caption{Light curves of Lu et al. (1988) in the $B$ and $V$ bandpasses. The continuous curves represent the solutions 
 obtained with the model parameters listed in the third column of Table 3. }
 \label{Fig4}
\end{figure}

\begin{figure}
 \includegraphics[]{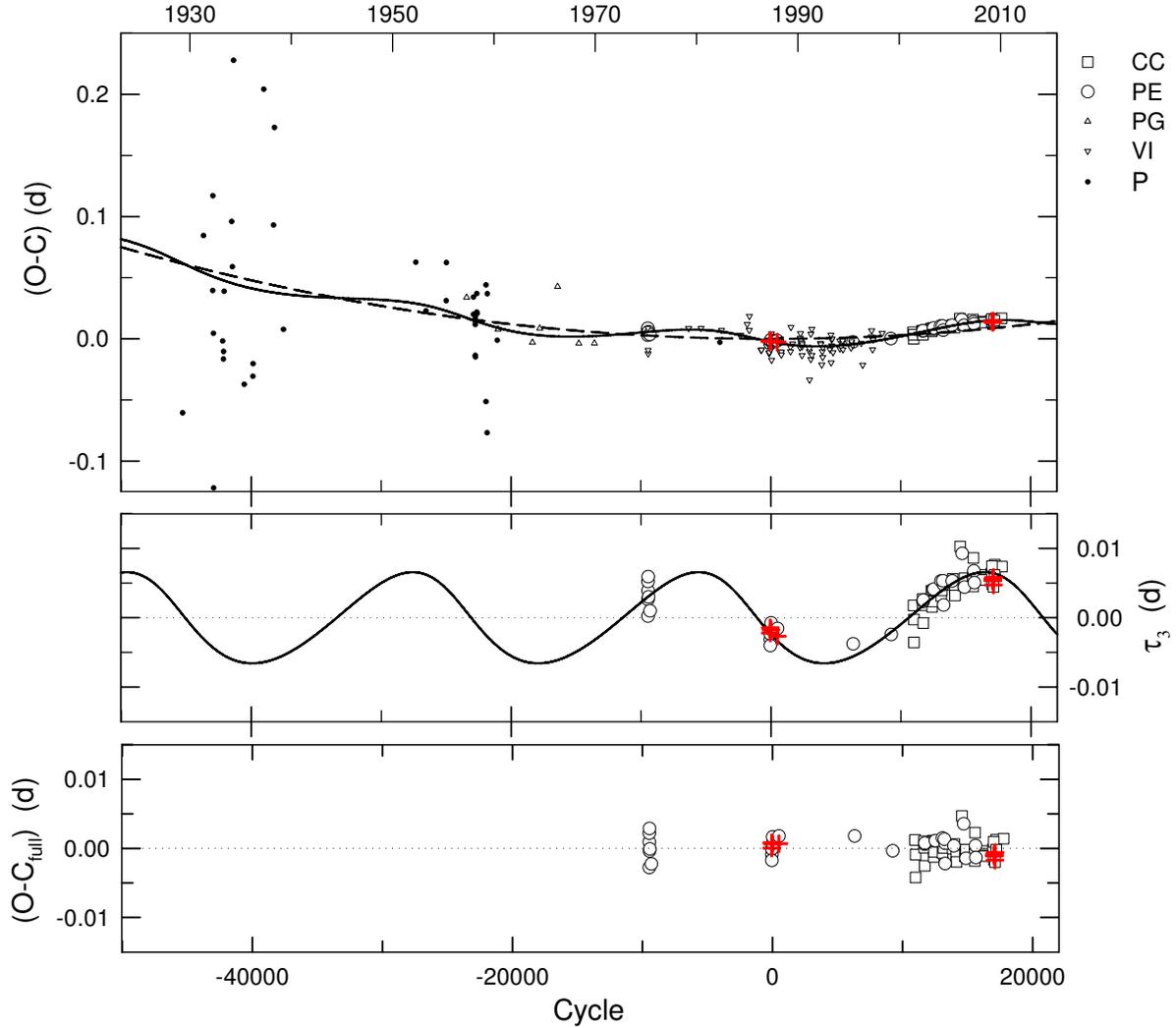}
 \caption{$O$--$C$ diagram of AA UMa. In the top panel, constructed with the linear terms of the quadratic {\it plus} 
 LTT ephemeris, the continuous curve and the dashed, parabolic one represent the full contribution and the quadratic term 
 of the equation, respectively. The middle panel represents the LTT orbit, and the bottom panel the PE and CCD residuals 
 from the complete ephemeris. CC, PE, PG, VI, and P denote CCD, photoelectric, photographic, visual, and 
 photographic plate minima, respectively. Plus symbols refer to new minimum timings obtained with the W-D code. }
 \label{Fig5}
\end{figure}

\clearpage
\begin{deluxetable}{crcrcr}
\tablewidth{0pt} 
\tablecaption{CCD photometric observations of AA UMa.}
\tablehead{
\colhead{HJD} & \colhead{$\Delta B$} & \colhead{HJD} & \colhead{$\Delta V$} & \colhead{HJD} & \colhead{$\Delta R$} 
}
\startdata
2,454,840.23153 &    0.351  &  2,454,840.22922  &  0.560  &  2,454,840.22778  &  0.685   \\
2,454,840.23741 &    0.373  &  2,454,840.23510  &  0.594  &  2,454,840.23365  &  0.727   \\
2,454,840.24305 &    0.352  &  2,454,840.24095  &  0.593  &  2,454,840.23952  &  0.735   \\
2,454,840.24820 &    0.308  &  2,454,840.24627  &  0.553  &  2,454,840.24492  &  0.708   \\
2,454,840.25327 &    0.250  &  2,454,840.25132  &  0.503  &  2,454,840.24998  &  0.663   \\
2,454,840.25834 &    0.184  &  2,454,840.25639  &  0.447  &  2,454,840.25505  &  0.605   \\
2,454,840.26341 &    0.120  &  2,454,840.26146  &  0.386  &  2,454,840.26012  &  0.551   \\
2,454,840.26848 &    0.058  &  2,454,840.26653  &  0.328  &  2,454,840.26519  &  0.486   \\
2,454,840.27355 &    0.004  &  2,454,840.27161  &  0.271  &  2,454,840.27027  &  0.435   \\
2,454,840.27880 & $-$0.045  &  2,454,840.27675  &  0.223  &  2,454,840.27536  &  0.387   \\
\enddata
\tablecomments{This table is available in its entirety in machine-readable and Virtual Observatory (VO) forms 
in the online journal. A portion is shown here for guidance regarding its form and content.}
\end{deluxetable}

\begin{deluxetable}{lcccccc}
\tabletypesize{\small}
\tablewidth{0pt}
\tablecaption{Photometric solutions of AA UMa.}
\tablehead{
\colhead{Parameter}                    & \colhead{Without Spot}  & \multicolumn{4}{c}{Spot Model$\rm ^a$}  \\ [1.0mm] \cline{3-6} \\ [-2.0ex]
                                       &                  & \colhead{Cool 1} & \colhead{Hot 1}  & \colhead{Cool 2} & \colhead{Hot 2}                    
}                                                                                                                                                         
\startdata                                                                                                                                                
$T_0$ (HJD)$^b$                        &  867.15532(11)   &  867.15463(6)    &  867.15522(6)    &  867.15463(6)    &  867.15522(6)    \\                  
$P$ (d)                                &  0.4681347(29)   &  0.4681348(16)   &  0.4681356(16)   &  0.4681348(16)   &  0.4681350(16)   \\
$q$ (=$m\rm_2/m\rm_1$)                 &  1.8138(95)      &  1.8157(88)      &  1.8090(63)      &  1.8157(88)      &  1.8102(10)      \\                  
$i$ (deg)                              &  79.20(13)       &  79.68(8)        &  78.57(8)        &  79.61(8)        &  78.57(7)        \\                  
$T_1$ (K)                              &  5964(9)         &  5963(5)         &  5971(5)         &  5964(5)         &  5971(5)         \\                  
$\Omega_1$=$\Omega_2$                  &  4.906(14)       &  4.904(12)       &  4.946(11)       &  4.904(12)       &  4.947(35)       \\                  
$f$ (\%)                               &  13.5            &  14.3            &  5.5             &  14.3            &  5.8             \\                  
$A_1$=$A_2$                            &  0.50            &  0.50            &  0.50            &  0.50            &  0.50            \\
$g_1$=$g_2$                            &  0.32            &  0.32            &  0.32            &  0.32            &  0.32            \\
$X$, $Y$                               &  0.644, 0.219    &  0.644, 0.219    &  0.644, 0.219    &  0.644, 0.219    &  0.644, 0.219    \\                  
$x_{B}$, $y_{B}$                       &  0.829, 0.174    &  0.829, 0.174    &  0.828, 0.175    &  0.829, 0.174    &  0.828, 0.175    \\                  
$x_{V}$, $y_{V}$                       &  0.746, 0.252    &  0.746, 0.252    &  0.746, 0.253    &  0.746, 0.252    &  0.746, 0.253    \\                  
$x_{R}$, $y_{R}$                       &  0.654, 0.266    &  0.654, 0.266    &  0.653, 0.266    &  0.654, 0.266    &  0.653, 0.266    \\                  
$L_1$/($L_{1}$+$L_{2}$){$_{B}$}        &  0.3796(22)      &  0.3793(14)      &  0.3803(13)      &  0.3796(14)      &  0.3802(12)      \\                  
$L_1$/($L_{1}$+$L_{2}$){$_{V}$}        &  0.3767(17)      &  0.3765(12)      &  0.3769(11)      &  0.3767(13)      &  0.3768(10)      \\                  
$L_1$/($L_{1}$+$L_{2}$){$_{R}$}        &  0.3751(15)      &  0.3749(11)      &  0.3751(9)       &  0.3751(11)      &  0.3750(8)       \\                                                                                                                                                        
$r_1$ (pole)                           &  0.3148(16)      &  0.3152(14)      &  0.3107(12)      &  0.3152(14)      &  0.3107(3)       \\                  
$r_1$ (side)                           &  0.3299(20)      &  0.3304(17)      &  0.3249(14)      &  0.3304(17)      &  0.3250(4)       \\                  
$r_1$ (back)                           &  0.3671(33)      &  0.3679(28)      &  0.3591(23)      &  0.3679(28)      &  0.3592(6)       \\                  
$r_1$ (volume)$^c$                     &  0.3393          &  0.3399          &  0.3333          &  0.3399          &  0.3334          \\                  
$r_2$ (pole)                           &  0.4132(14)      &  0.4138(12)      &  0.4088(11)      &  0.4138(11)      &  0.4090(3)       \\                  
$r_2$ (side)                           &  0.4396(18)      &  0.4403(15)      &  0.4339(14)      &  0.4403(15)      &  0.4341(4)       \\                  
$r_2$ (back)                           &  0.4711(25)      &  0.4720(21)      &  0.4636(19)      &  0.4720(21)      &  0.4638(5)       \\                  
$r_2$ (volume)$^c$                     &  0.4432          &  0.4439          &  0.4370          &  0.4439          &  0.4373          \\ [1.0mm]          
\multicolumn{6}{l}{Spot parameters:}                                                                                                  \\                  
Colatitude$_1$ (deg)                   &  \dots           &  72.44(70)       &  69.28(26)       &  \dots           &  \dots           \\                  
Longitude$_1$ (deg)                    &  \dots           &  276.47(86)      &  100.69(48)      &  \dots           &  \dots           \\                  
Radius$_1$ (deg)                       &  \dots           &  20.72(20)       &  19.87(12)       &  \dots           &  \dots           \\                  
$T$$\rm _{spot1}$/$T$$\rm _{local1}$   &  \dots           &  0.7566(88)      &  1.1464(24)      &  \dots           &  \dots           \\                  
Colatitude$_2$ (deg)                   &  \dots           &  \dots           &  \dots           &  73.12(26)       &  69.30(84)       \\                  
Longitude$_2$ (deg)                    &  \dots           &  \dots           &  \dots           &  98.15(71)       &  276.46(96)      \\                  
Radius$_2$ (deg)                       &  \dots           &  \dots           &  \dots           &  20.56(22)       &  20.80(54)       \\                  
$T$$\rm _{spot2}$/$T$$\rm _{local2}$   &  \dots           &  \dots           &  \dots           &  0.8776(30)      &  1.0854(17)      \\ [1.0mm]          
$\Sigma W(O-C)^2$                      &  0.0215          &  0.0119          &  0.0122          &  0.0119          &  0.0120                         
\enddata
\tablenotetext{a}{Cool 1: a cool spot on the primary; Hot 1: a hot spot on the primary; Cool 2: a cool spot on the secondary; 
 Hot 2: a hot spot on the secondary. }
\tablenotetext{b}{HJD 2,454,000 is suppressed.}
\tablenotetext{c}{Mean volume radius.}
\end{deluxetable}

\begin{deluxetable}{lcc}
\tablewidth{0pt}
\tablecaption{AA UMa parameters obtained from the Lu et al. light curves.}
\tablehead{
\colhead{Parameter}                    & \colhead{Cool 1}     & \colhead{Cool 2}
}                                                          
\startdata                                                 
$T_0$ (HJD)                            &  2,446,857.02450(8)  &  2,446,857.02450(8)   \\        
$P$ (d)                                &  0.46812501(43)      &  0.46812501(42)       \\        
$q$ (=$m\rm_2/m\rm_1$)                 &  1.819(14)           &  1.8157(88)           \\        
$i$ (deg)                              &  80.42(10)           &  80.42(8)             \\        
$T_1$ (K)                              &  5963(5)             &  5963(5)              \\        
$\Omega_1$=$\Omega_2$                  &  4.907(18)           &  4.909(20)            \\        
$f$ (\%)                               &  14.6                &  14.8                 \\        
$L_1$/($L_{1}$+$L_{2}$){$_{B}$}        &  0.3790(18)          &  0.3786(17)           \\        
$L_1$/($L_{1}$+$L_{2}$){$_{V}$}        &  0.3761(17)          &  0.3758(17)           \\        
$r_1$ (pole)                           &  0.3152(21)          &  0.3153(23)           \\        
$r_1$ (side)                           &  0.3304(26)          &  0.3305(29)           \\        
$r_1$ (back)                           &  0.3680(44)          &  0.3682(49)           \\        
$r_1$ (volume)                         &  0.3400              &  0.3401               \\        
$r_2$ (pole)                           &  0.4132(14)          &  0.4144(20)           \\        
$r_2$ (side)                           &  0.4396(18)          &  0.4410(26)           \\        
$r_2$ (back)                           &  0.4711(25)          &  0.4728(36)           \\        
$r_2$ (volume)                         &  0.4443              &  0.4447               \\ [1.0mm]
\multicolumn{3}{l}{Spot parameters:}                                                  \\        
Colatitude (deg)                       &  83.22(82)           &  82.82(45)            \\        
Longitude (deg)                        &  252.74(97)          &  76.21(28)            \\        
Radius (deg)                           &  21.65(32)           &  18.25(13)            \\        
$T$$\rm _{spot}$/$T$$\rm _{local}$     &  0.782(21)           &  0.822(8)             \\ [1.0mm]
$\Sigma W(O-C)^2$                      &  0.0169              &  0.0168               
\enddata
\end{deluxetable}

\begin{deluxetable}{lrrcl}
\tabletypesize{\small} 
\tablewidth{0pt}
\tablecaption{Photoelectric and CCD timings of minimum light for AA UMa.}
\tablehead{
\colhead{HJD} & \colhead{Epoch} & \colhead{$O$--$C_{\rm full}$} & \colhead{Min} & References \\
\colhead{(2,400,000+)} & & & & }                                                               
\startdata                                                                                     
42,450.5570  &  $-$9473.0  &     0.00091  &  I   &  Meinunger (1976)              \\  
42,451.2555  &  $-$9471.5  &  $-$0.00278  &  II  &  Meinunger (1976)              \\  
42,452.4308  &  $-$9469.0  &     0.00220  &  I   &  Meinunger (1976)              \\  
42,452.6623  &  $-$9468.5  &  $-$0.00037  &  II  &  Meinunger (1976)              \\  
42,454.3040  &  $-$9465.0  &     0.00289  &  I   &  Meinunger (1976)              \\  
42,472.5580  &  $-$9426.0  &  $-$0.00009  &  I   &  Meinunger (1976)              \\  
42,524.5180  &  $-$9315.0  &  $-$0.00224  &  I   &  Meinunger (1976)              \\  
46,857.2571  &    $-$59.5  &  $-$0.00089  &  II  &  Lu et al. (1988)              \\  
46,859.1304  &    $-$55.5  &  $-$0.00009  &  II  &  Lu et al. (1988)              \\  
46,860.0650  &    $-$53.5  &  $-$0.00174  &  II  &  Lu et al. (1988)              \\  
46,885.1121  &      0.0    &     0.00069  &  I   &  Lu et al. (1988)              \\  
46,886.0493  &      2.0    &     0.00165  &  I   &  Lu et al. (1988)              \\  
47,118.2393  &    498.0    &     0.00180  &  I   &  Lu et al. (1988)              \\  
49,840.3945  &   6313.0    &     0.00180  &  I   &  Diethelm (1995)               \\  
51,209.4335  &   9237.5    &  $-$0.00035  &  II  &  Agerer \& H\"ubscher (2000)   \\  
52,016.0209  &  10960.5    &     0.00121  &  II  &  Nagai (2002)                  \\  
52,032.400   &  10995.5    &  $-$0.00420  &  II  &  Diethelm (2001)               \\  
52,032.4033  &  10995.5    &  $-$0.00090  &  II  &  Agerer \& H\"ubscher (2002)   \\  
52,339.9659  &  11652.5    &     0.00103  &  II  &  Nagai (2003)                  \\  
52,344.4122  &  11662.0    &     0.00011  &  I   &  Diethelm (2002)               \\  
52,361.4994  &  11698.5    &     0.00061  &  II  &  Agerer \& H\"ubscher (2003)   \\  
52,367.8160  &  11712.0    &  $-$0.00253  &  I   &  Pribulla et al. (2002)        \\  
52,368.5215  &  11713.5    &     0.00077  &  II  &  Agerer \& H\"ubscher (2003)   \\  
52,658.5277  &  12333.0    &     0.00114  &  I   &  Diethelm (2003)               \\  
52,688.0197  &  12396.0    &     0.00102  &  I   &  Nagai (2004)                  \\  
52,688.2515  &  12396.5    &  $-$0.00124  &  II  &  Nagai (2004)                  \\  
52,698.3170  &  12418.0    &  $-$0.00051  &  I   &  Brat (2007)                   \\  
52,742.5568  &  12512.5    &     0.00111  &  II  &  H\"ubscher (2005)             \\  
52,745.3655  &  12518.5    &     0.00104  &  II  &  Pribulla et al. (2005)        \\  
53,003.5390  &  13070.0    &     0.00150  &  I   &  H\"ubscher (2005)             \\  
53,010.3247  &  13084.5    &  $-$0.00067  &  II  &  Nagai (2005)                  \\  
53,040.7538  &  13149.5    &     0.00006  &  II  &  Nelson (2005)                 \\  
53,069.5450  &  13211.0    &     0.00134  &  I   &  H\"ubscher et al. (2005)      \\  
53,095.5226  &  13266.5    &  $-$0.00221  &  II  &  H\"ubscher et al. (2005)      \\  
53,096.4618  &  13268.5    &     0.00074  &  II  &  H\"ubscher (2005)             \\  
53,363.2944  &  13838.5    &  $-$0.00003  &  II  &  Nagai (2005)                  \\  
53,410.5758  &  13939.5    &     0.00038  &  II  &  H\"ubscher et al. (2005)      \\  
53,433.2803  &  13988.0    &     0.00063  &  I   &  Kim et al. (2006)             \\  
53,503.0298  &  14137.0    &  $-$0.00102  &  I   &  Kim et al. (2006)             \\  
53,510.0508  &  14152.0    &  $-$0.00195  &  I   &  Kim et al. (2006)             \\  
53,700.3517  &  14558.5    &     0.00471  &  II  &  Nagai (2006)                  \\  
53,765.4204  &  14697.5    &     0.00356  &  II  &  H\"ubscher (2007)             \\  
53,814.3351  &  14802.0    &  $-$0.00116  &  I   &  H\"ubscher et al. (2006)      \\  
53,846.4016  &  14870.5    &  $-$0.00145  &  II  &  Diethelm (2006)               \\  
53,846.4029  &  14870.5    &  $-$0.00015  &  II  &  H\"ubscher et al. (2006)      \\  
54,154.8989  &  15529.5    &  $-$0.00065  &  II  &  Nelson (2008)                 \\  
54,159.5805  &  15539.5    &  $-$0.00033  &  II  &  Dvorak (2008)                 \\  
54,167.5372  &  15556.5    &  $-$0.00181  &  II  &  H\"ubscher et al. (2009)      \\  
54,173.3929  &  15569.0    &     0.00229  &  I   &  Diethelm (2007)               \\  
54,186.4986  &  15597.0    &     0.00041  &  I   &  H\"ubscher (2007)             \\  
54,206.3923  &  15639.5    &  $-$0.00134  &  II  &  H\"ubscher (2007)             \\  
54,469.4804  &  16201.5    &  $-$0.00111  &  II  &  Parimucha et al. (2009)       \\  
54,498.7387  &  16264.0    &  $-$0.00079  &  I   &  Nelson (2009)                 \\  
54,521.4433  &  16312.5    &  $-$0.00038  &  II  &  H\"ubscher et al. (2009)      \\  
54,540.4018  &  16353.0    &  $-$0.00105  &  I   &  Brat et al. (2008)            \\  
54,816.8320  &  16943.5    &  $-$0.00012  &  II  &  Nelson (2009)                 \\  
54,845.857   &  17005.5    &     0.00099  &  II  &  Diethelm (2009)               \\  
54,854.7488  &  17024.5    &  $-$0.00163  &  II  &  Dvorak (2010)                 \\ 
54,867.1559  &  17051.0    &     0.00006  &  I   &  This paper (SOAO)             \\  
54,869.0285  &  17055.0    &     0.00018  &  I   &  This paper (SOAO)             \\  
54,874.1778  &  17066.0    &     0.00013  &  I   &  This paper (SOAO)             \\  
54,895.0074  &  17110.5    &  $-$0.00197  &  II  &  This paper (SOAO)             \\  
54,897.1166  &  17115.0    &     0.00070  &  I   &  This paper (SOAO)             \\  
54,937.3747  &  17201.0    &  $-$0.00018  &  I   &  Parimucha et al. (2009)       \\  
54,941.3544  &  17209.5    &     0.00044  &  II  &  H\"ubscher et al. (2010)      \\  
54,941.5893  &  17210.0    &     0.00128  &  I   &  H\"ubscher et al. (2010)      \\  
55,201.8680  &  17766.0    &     0.00142  &  I   &  Diethelm (2010)               \\ 
\enddata
\end{deluxetable}

\begin{deluxetable}{lcc}
\tablewidth{0pt}
\tablecaption{Parameters for the quadratic {\it plus} LTT ephemeris of AA UMa. }
\tablehead{
\colhead{Parameter}            &  \colhead{Values}            &  \colhead{Unit}
}
\startdata
$T_0$                          &  2,446,885.11379(87)         &  HJD                   \\
$P$                            &  0.468126612(66)             &  d                     \\
$A$                            &  2.99(44)$\times 10^{-11}$   &  d                     \\
$a_{12}\sin i_{3}$             &  1.16(21)                    &  AU                    \\
$\omega$                       &  154(10)                     &  deg                   \\
$e$                            &  0.22(19)                    &                        \\
$n  $                          &  0.0349(20)                  &  deg d$^{-1}$          \\
$T$                            &  2,445,788(311)              &  HJD                   \\
$P_{3}$                        &  28.2(1.6)                   &  yr                    \\
$K$                            &  0.0066(12)                  &  d                     \\
$f(M_{3})$                     &  0.00196(38)                 &  $M_\odot$             \\
$M_3$ ($i_{3}$=90 deg)$\rm^a$  &  0.25                        &  $M_\odot$             \\
$M_3$ ($i_{3}$=60 deg)$\rm^a$  &  0.29                        &  $M_\odot$             \\
$M_3$ ($i_{3}$=30 deg)$\rm^a$  &  0.52                        &  $M_\odot$             \\
$dP$/$dt$                      &  4.68(69)$\times 10^{-8}$    &  d yr$^{-1}$           \\
$dM_2$/$dt$                    &  6.63$\times 10^{-8}$        &  $M_\odot$ yr$^{-1}$   \\
\enddata
\tablenotetext{a}{Masses of the hypothetical third body for different values of $i_{3}$.}
\end{deluxetable}

\begin{deluxetable}{ccccccl}
\tablewidth{0pt}
\tablecaption{Minimum timings determined by the W-D code from individual eclipses of AA UMa.}
\tablehead{
\colhead{Observed$\rm^{a,b}$} & \colhead{W-D$\rm^{b}$} & \colhead{Error$\rm^{c}$} & \colhead{Difference$\rm^{d}$} & \colhead{Filter} & \colhead{Min} & References
}
\startdata
46,857.2571  &  46,857.25805  &  $\pm$0.00016  &  $-$0.00095  &  $BV$   &  II  &  Lu et al.      \\
46,859.1304  &  46,859.13131  &  $\pm$0.00012  &  $-$0.00091  &  $BV$   &  II  &  Lu et al.      \\
46,885.1121  &  46,885.11205  &  $\pm$0.00011  &  $+$0.00005  &  $BV$   &  I   &  Lu et al.      \\
47,118.2393  &  47,118.23819  &  $\pm$0.00051  &  $+$0.00111  &  $BV$   &  I   &  Lu et al.      \\
54,867.1559  &  54,867.15480  &  $\pm$0.00011  &  $+$0.00110  &  $BVR$  &  I   &  This article   \\
54,869.0285  &  54,869.02745  &  $\pm$0.00009  &  $+$0.00105  &  $BVR$  &  I   &  This article   \\
54,874.1778  &  54,874.17662  &  $\pm$0.00010  &  $+$0.00118  &  $BVR$  &  I   &  This article   \\
54,895.0074  &  54,895.00767  &  $\pm$0.00016  &  $-$0.00027  &  $BVR$  &  II  &  This article   \\
54,897.1166  &  54,897.11534  &  $\pm$0.00020  &  $+$0.00126  &  $BVR$  &  I   &  This article   \\
\enddata
\tablenotetext{a}{cf. Table 4.}
\tablenotetext{b}{HJD 2,400,000 is suppressed.}
\tablenotetext{c}{Uncertainties yielded by the W-D code.}
\tablenotetext{d}{Differences between columns (1) and (2).}
\end{deluxetable}

\begin{deluxetable}{lcccc}
\tablewidth{0pt} 
\tablecaption{Astrophysical parameters for AA UMa.}
\tablehead{
\colhead{Parameter}      & \colhead{Wang \& Lu}  & \colhead{This Paper}  & \colhead{Unit}     
}                                                                                             
\startdata
$a$                      &  3.39                 &  3.44$\pm$0.03                 &  $R_\odot$       \\
$V_0$                    &  $-$35.6              &  $-$35.3$\pm$1.5               &  km s$^{-1}$     \\
$M_1$                    &  0.85                 &  0.89$\pm$0.02                 &  $M_\odot$       \\
$M_2$                    &  1.55                 &  1.61$\pm$0.03                 &  $M_\odot$       \\
$R_1$                    &  1.14                 &  1.17$\pm$0.02                 &  $R_\odot$       \\
$R_2$                    &  1.50                 &  1.53$\pm$0.02                 &  $R_\odot$       \\
$L_1$                    &  1.55                 &  1.55$\pm$0.17                 &  $L_\odot$       \\
$L_2$                    &  2.48                 &  2.57$\pm$0.27                 &  $L_\odot$       \\
$M_{\rm bol1}$           &                       &  $+$4.21$\pm$0.12              &  mag             \\
$M_{\rm bol2}$           &                       &  $+$3.67$\pm$0.11              &  mag             \\
BC$_1$                   &                       &  $-$0.05                       &  mag             \\
BC$_2$                   &                       &  $-$0.06                       &  mag             \\
$M_{V1}$                 &                       &  $+$4.26$\pm$0.12              &  mag             \\
$M_{V2}$                 &                       &  $+$3.73$\pm$0.11              &  mag             \\
Distance                 &                       &  351$\pm$19                    &  pc              \\
\enddata
\end{deluxetable}

\begin{deluxetable}{lccc}
\tablewidth{0pt}
\tablecaption{Applegate parameters for possible magnetic activity of AA UMa.}
\tablehead{
\colhead{Parameter}       & \colhead{Primary}      & \colhead{Secondary}     & \colhead{Unit}
}
\startdata
$\Delta P$                & 0.1628                 &  0.1628                 &  s                   \\
$\Delta P/P$              & $4.03\times10^{-6}$    &  $4.03\times10^{-6}$    &                      \\
$\Delta Q$                & ${4.54\times10^{49}}$  &  ${8.22\times10^{49}}$  &  g cm$^2$            \\
$\Delta J$                & ${1.92\times10^{47}}$  &  ${2.81\times10^{47}}$  &  g cm$^{2}$ s$^{-1}$ \\
$I_{\rm s}$               & ${7.83\times10^{53}}$  &  ${2.42\times10^{54}}$  &  g cm$^{2}$          \\
$\Delta \Omega$           & ${2.45\times10^{-7}}$  &  ${1.16\times10^{-7}}$  &  s$^{-1}$            \\
$\Delta \Omega / \Omega$  & ${1.58\times10^{-3}}$  &  ${7.46\times10^{-4}}$  &                      \\
$\Delta E$                & ${9.40\times10^{40}}$  &  ${6.51\times10^{40}}$  &  erg                 \\
$\Delta L_{\rm rms}$      & ${3.32\times10^{32}}$  &  ${2.29\times10^{32}}$  &  erg s$^{-1}$        \\
                          & 0.085                  &  0.059                  &  $L_\odot$           \\
                          & 0.055                  &  0.023                  &  $L_{1,2}$           \\
$\Delta m_{\rm rms}$      & $\pm$0.022             &  $\pm$0.015             &  mag                 \\
$B$                       & 8,878                  &  7,181                  &  G
\enddata
\end{deluxetable}

\end{document}